\documentclass[12pt]{article}
\usepackage{layout}
\usepackage[margin=2.5cm]{geometry}
\usepackage{graphicx} 
\usepackage{amsmath}
\usepackage{graphicx}
\usepackage[colorinlistoftodos]{todonotes}
\usepackage[colorlinks=true, allcolors=blue]{hyperref}
\usepackage{indentfirst}
\usepackage{setspace}
\usepackage{caption}
\usepackage{subcaption}
\usepackage[autostyle]{csquotes}

\begin{document}
\singlespacing
\begin{titlepage}
	{\centering
        \vspace*{5cm}
	{\huge\bfseries Dynamic Exclusion of Low-Fidelity Data in Bayesian Optimization for Autonomous Beamline Alignment\par}
		\vspace{1cm}
	{\Large Megha Ragini Narayanan}\\
 {\large School of Computer Science \\ Carnegie Mellon University \\ Pittsburgh, PA 15213\par}
    \vspace{0.5cm}
	{\Large Thomas W. Morris}\\
 {\large
 National Synchrotron Light Source II\\ Brookhaven National Laboratory\\ Upton, NY, 11973\\
 Department of Physics \\ Yale University \\  New Haven, CT 06511\par}
    \vspace{1cm}
	{\large August 8, 2024 \par}}
\end{titlepage}

\doublespacing

\begin{center}
\textbf{Abstract}
\end{center}

\begin{changemargin}{1cm}{1cm} 

Aligning beamlines at synchrotron light sources is a high-dimensional, expensive-to-sample optimization problem, as beams are focused using a series of dynamic optical components.  Bayesian optimization is an efficient machine learning approach to finding global optima of beam quality, but the model can easily be impaired by faulty data points caused by the beam going off the edge of the sensor or by background noise.  This study, conducted at the National Synchrotron Light Source II (NSLS-II) facility at Brookhaven National Laboratory (BNL), is an investigation of methods to identify untrustworthy readings of beam quality and discourage the optimization model from seeking out points likely to yield low-fidelity beams.  The approaches explored include dynamic pruning using loss analysis of size and position models and a lengthscale-based genetic algorithm to determine which points to include in the model for optimal fit.  Each method successfully classified high and low fidelity points.  This research advances BNL’s mission to tackle our nation’s energy challenges by providing scientists at all beamlines with access to higher quality beams, and faster convergence to these optima for their experiments.  
\end{changemargin}

\setcounter{page}{2}

\section{Introduction}
\subsection{Background}

\noindent The invaluable work performed by scientists at the National Synchrotron Light Source II (NSLS-II) is made possible by the high-quality beams produced by each of their beamlines. These beamlines are focused using a series of precision optical components, including crystals, mirrors, and aperture shutters. Each component is dynamic, with multiple degrees of freedom, and can be adjusted using precision motors within the beamline to optimize the final beam image. Every time a beamline is used, these components must be precisely reconfigured to achieve optimal beam quality—a beam that is both small in size and bright in intensity.

However, the sensitivity of these components to small environmental changes, such as temperature fluctuations, vibrations, and mechanical drifts, adds significant complexity to the alignment process. Additionally, as the scientific demands grow, beamlines are becoming increasingly sophisticated, with more components and greater degrees of freedom. This escalation in complexity makes manual alignment not only time-consuming but also prone to human error.
\begin{figure}[h!]
    \centering
    \includegraphics[width=1\linewidth]{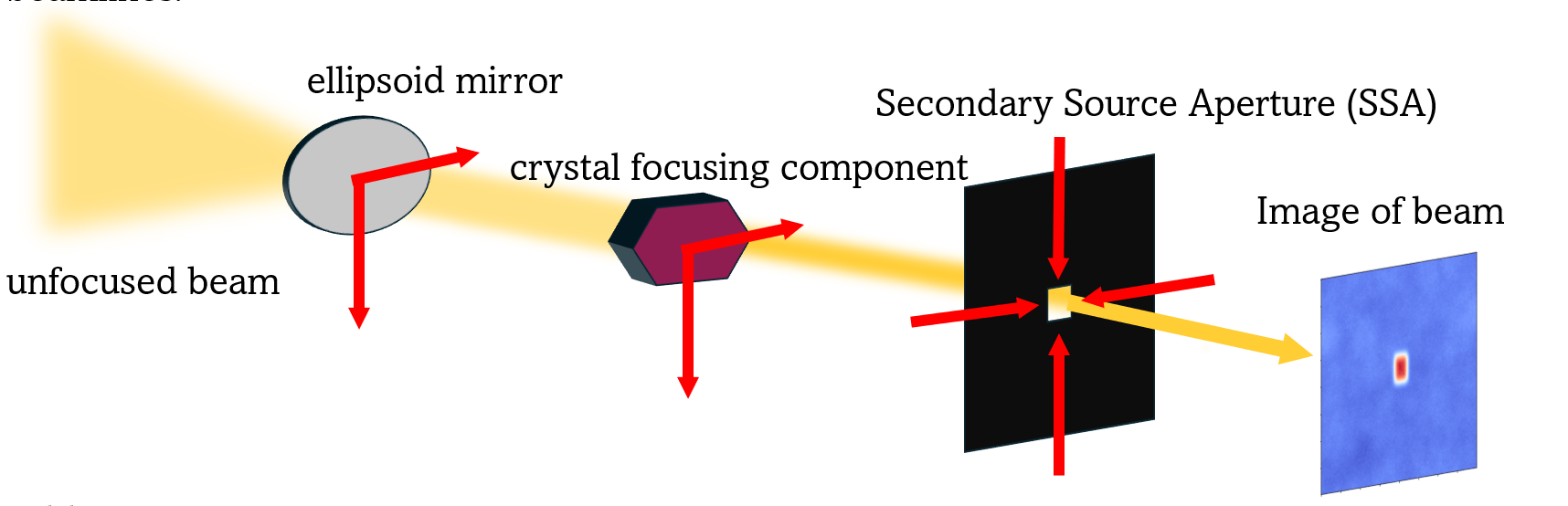}
    \caption{Example of extremely simple beamline focusing setup, with degrees of freedom shown in red}
    \label{fig:enter-label}
\end{figure}

Traditionally, the alignment of beamlines has been performed manually. However, the challenges posed by the intricate and sensitive nature of modern beamlines have highlighted the need for more efficient and reliable methods. In this context, robotic assistance for autonomous beamline alignment presents a highly promising solution. Autonomous methods, leveraging machine learning and real-time feedback systems, can perform adjustments with greater precision and consistency. They have the potential to converge on optimal settings faster and more accurately than manual methods, leading to significant improvements in both beam quality and experimental efficiency.

Implementing autonomous alignment systems could revolutionize the operation of synchrotron facilities, enhancing the reproducibility of experiments and maximizing the scientific output. By reducing the dependency on manual intervention, scientists can focus more on their research objectives, confident in the knowledge that the beamlines are consistently delivering the highest quality beams possible.

\subsection{Autonomous Methods}

\noindent Currently, we are using Gaussian Processes and Bayesian Optimization to find global optima of beam quality, tools which are uniquely suited for expensive-to-sample problems with multiple objectives such as this one.$^{1}$  At a high level, we are looking for the input $x$ which leads to the maximum value of the true function $f(x)$, which relates our inputs in terms of component positions to outputs in terms of beam quality.  Given that we must sample as few points as possible, we treat this function as a stochastic process and use Bayesian inference to create a posterior distribution, which represents the probability a given function is the true function.  This posterior can be visualised as a mean with error bars, and given an input $x$, our posterior gives us a distribution of what $y$ is likely to be.  In each iteration of  Bayesian inference, we perform three main steps.  First, we create a posterior distribution $p(f)$ based on historical observations $(x, y)$, according to the equation below, where the likelihood $p(y\mid f, x)$ is the probability of observing outputs $y$ at inputs $x$ for a given function $f$, and the prior $p(f)$ is our knowledge of the likelihood of a function $f$ before we have observed any data. 

$$p(f\mid x, y) = \frac{p(y\mid f, x) p(f)}{p(y\mid x)}$$

This is generally done using Gaussian Process (GP) models which optimize hyperparameters of a kernel to estimate a covariance matrix and construct a posterior.  Second, we use an acquisition function such as Expected Improvement (EI) for single objectives, or Noisy Expected Hypervolume Improvement (NEHVI) for multiple objectives, which estimates how much better than the existing optima a given point will be, to determine the best points to sample. Finally, we sample these points and add them to the historical observations.

\subsection{Low-Fidelity Data}

\noindent The beam is optimized based on images from sensors within the beamline.  However, one key complication with autonomous alignment is that these sensors can be finnicky, and beam data noisy.  This leads to slower convergence from our model, as it can be impaired by these faulty data points caused e.g. by the beam going off the edge of the sensor or by background noise.  Thus, this study is an investigation of methods to identify untrustworthy readings of beam quality and discourage the optimization model from seeking out points likely to yield low-fidelity beams.  One elementary solution would be to set a flux cutoff for a point to be considered by our model, which would eliminate images where the beam is no longer on the sensor, or so unfocused that minimal light is hitting the sample.  However, different beamlines produce very different measures of flux, and these measurements are often impacted by other conditions which makes this cutoff hard to set.  Since our goal is to be able to apply this machine learning framework to alignment at many different beamlines, we present this study to investigate methods to remove these faulty data points in an online fashion, hopefully providing scientists at all beamlines with access to higher quality beams, and faster convergence to these optima for their experiments.     

\section{Methods and Results}

\noindent All of the models developed were originally tested on a simulated beam function, which takes 8 inputs, which model the motion of two elliptical mirrors and two Secondary Source Apertures (SSAs).  Each elliptical mirror has two degrees of freedom: vertical and horizontal.  Note that only the horizontal degrees of freedom affect the beam's width, and only the vertical ones affect the height.  A fairly high amount of correlated noise is added to the image produced by the simulated beam function. 

\subsection{Image Processing}

\noindent The first step in training models to optimize beam quality is processing the beam images into meaningful outputs which represent its quality.  Because of the high level of noise in the image, we first perform Singular Value Decomposition on the images, and then reconstruct them from only the most significant singular values, a linear algebra process which reduces images to their most significant features, as seen below.

\begin{figure}[h!]
\centering
\begin{subfigure}{.5\textwidth}
  \centering
  \includegraphics[width=.6\linewidth]{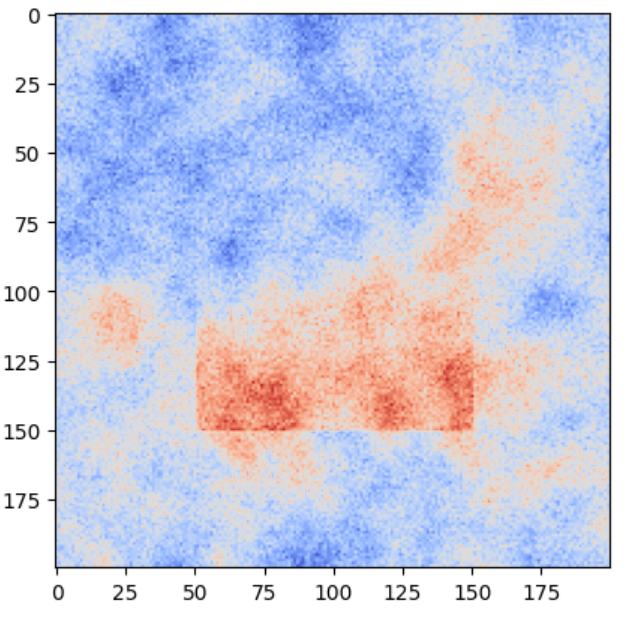}
  \caption{Example of raw beam image}
  \label{fig:sub1}
\end{subfigure}%
\begin{subfigure}{.5\textwidth}
  \centering
  \includegraphics[width=.6\linewidth]{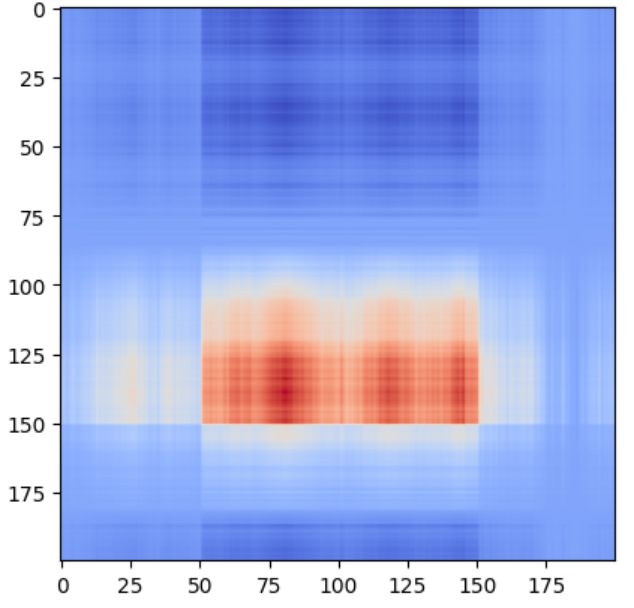}
  \caption{Same beam image, reconstructed after SVD.}
  \label{fig:sub2}
\end{subfigure}
\caption{Noise Reduction in Image Processing}
\label{fig:test}
\end{figure}

In the reconstructed image, the edges of the beam are identified, and the pixels within the beam are summed to yield the objective flux, representing brightness of the beam.  From the beam boundaries, the image processing function also outputs the height, width, x-position, and y-position of the beam. 

\subsection{Exclusion Based on Loss from Size and Position Models}

\noindent Due to the nature of the input variables, we expect each input to have a fairly smooth result on the image of the final beam, meaning if we move one motor, the beam should change in size or position in a predictable way.  However, for the \enquote{junk} beam images that are mostly noise, these transitions are not smooth, and we expect what the image digestion function finds to be the \enquote{center} of the beam to jump around.  As we can see below, as we change just 1 input, the beam moves predictably until the image becomes purely noise. 
\begin{figure}[h!]
    \centering
    \includegraphics[width=1\linewidth]{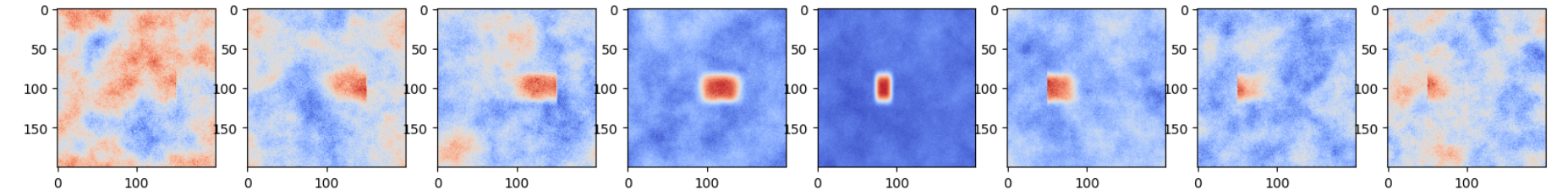}
    \caption{Beam images as we change just 1 variable along its entire domain}
    \label{fig:enter-label}
\end{figure}
If we add a few more data points to this, and train a model on the width and position of the beam, we see that the \enquote{good} beam images fit very well to the model, but the \enquote{junk} images that are mainly noise deviate from the model significantly.  Using this logic, we are able to prune points from the set of \enquote{good} beam readings by examining how much given points stray from the models.

We do this by first training models for the beams width, height, x-position, and y-position based on the data we assume to be \enquote{good}.  Then, we feed all of our data -- good and bad -- into these models and examine the marginal log likelihood for each point, which represents the probability of observing a input-output pair given the posterior of the model.  From this collection of loss values, we determine cutoff thresholds for each model, which determine how far off the model a point can be before it is labeled bad.

In these specific models, this threshold was chosen to be 1.2 multiplied by the 20th percentile of distances from the model, meaning a point had to be 1.2 times further from the model than the worst 20 percent of points to be pruned out of the data set.  If a point hit this threshold for any of the models-- width, height, x position, and y position, it was removed from the dataset.  One key feature of this approach is that the pruning was dynamic, so once a reading was deemed \enquote{bad} it could still be re-included in the dataset at a later iteration as the model changes, which accounts for complex function behavior.
\begin{figure}
    \centering
    \includegraphics[width=0.75\linewidth]{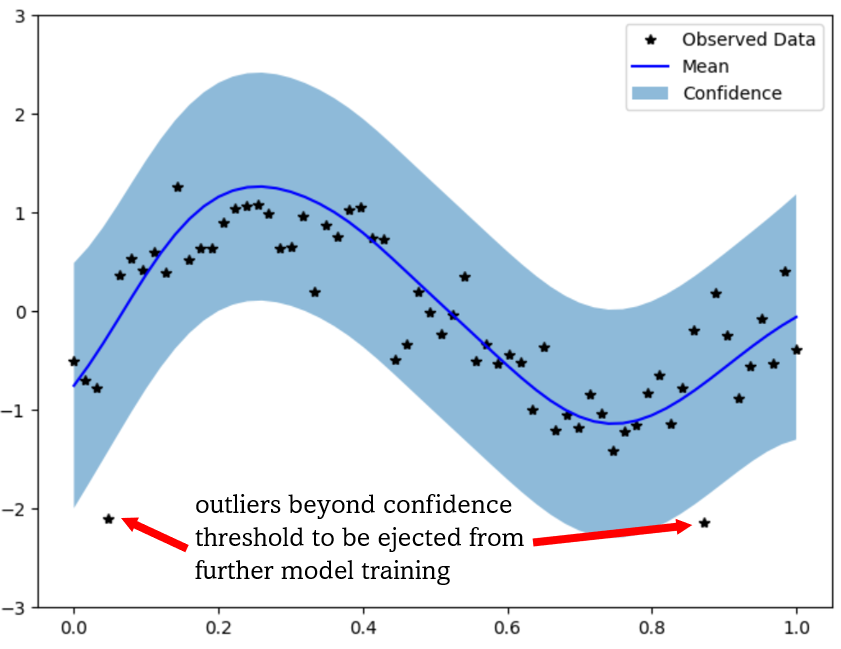}
    \caption{2-D example of points which would be pruned from dataset because they are too far from model posterior}
    \label{fig:enter-label}
\end{figure}

Originally, this pruning method was tested on a single-task GP optimizing for flux, with a constrained acquisition function which used traditional Expected Improvement, weighted by the posteriors of external width and height models, meaning the acquisition function favored points with smaller predicted sizes.  While this model was able to find optima, it was still slow to converge, as the constrained acquisition function meant the model was fitting to a much more complex function which related inputs to flux density (brightness per unit size), rather than understanding width, height, and flux as unique objectives which all had to be optimized in parallel.  The model was faster when we only trained the width and height models on the two dimensions they actually varied with, but in reality, we want to train all models on all dimensions as otherwise, some unexpected input having even a small effect on width or height could again throw off the model. Additionally, flux \enquote{levels out} at some point, which only made this fitting more difficult. 

Thus, we turned to utilizing true multi-objective Bayesian optimization of flux, width and height to reach optima faster.  Utilizing the dynamic pruning method outlined above, a multi-objective optimizing agent was able to find optima faster -- in just 30-40 points -- from our simulated function, working across four dimensions.  This pruning method worked alongside a trust domain restriction which excluded all points where the beam was percieved to have a width or height greater than half the size of the image.  Pictured below is one example of how the pruning method identified bad points.  The \enquote{True} or \enquote{False} above the image represents whether the point was deemed high-quality.  Further investigations into this algorithm may include fitting hyperparameters which weight how sensitive each of the sub-models are for loss confidence, allowing this algorithm to uniquely learn the specifics of each beamline.  

\begin{figure}[h!]
    \centering
    \includegraphics[width=0.8\linewidth]{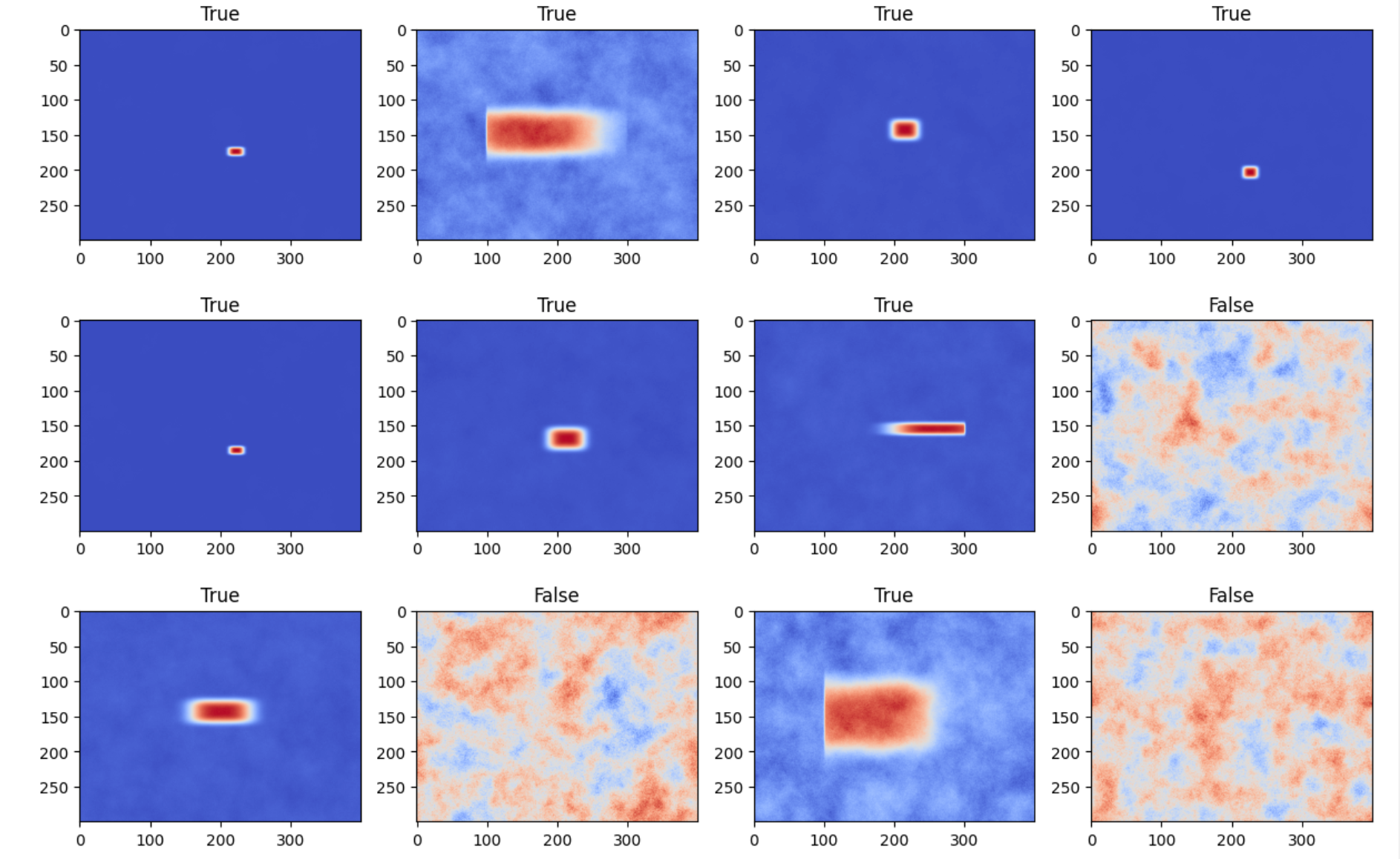}
    \caption{How dynamic pruning method classifies outputs}
    \label{fig:enter-label}
\end{figure}

\subsection{Genetic Algorithms}

\noindent Another way we can approach data pruning is as a binary classification problem, trying to find the optimal subset of datapoints to exclude for the best model training.  One method of binary classification is to employ a genetic algorithm, which takes inspiration from biology, simulating evolution to reach an optima.  In our genetic algorithm, we start with a population of \enquote{individuals}-- each of which has a unique genetic code.  The genetic code, in this case, is a binary string the length of our dataset, where the value at each index represents whether or not we are including the data point corresponding to that index.  From the initial population, the fitness of each individual is measured by constructing and fitting a model using the specific subset of data it encodes, and recording the lengthscale of this model. Similarly to the previous pruning method, we take advantage of the variety of beam image statistics available, and fit models to width, height, and position.  We factor the lengthscales of all of these models into our calculation of each individual's fitness. 

 The lengthscale is a hyperparameter of a model which represents how closely related we expect input points a certain distance away to be, which essentially tells us how \enquote{spiky} the fit posterior distributionwill be.  Because we expect our inputs to have a fairly smooth effect on beam quality, we consider smooth functions to be better fit, so we favor models with large lengthscales.  We also want to make sure that a model is fitting to a large enough number of points, so we also factor in the size of the dataset to fitness.  

 Once the fitness has been determined, the fit individuals are chosen to be parents, and offspring are formed by combining the genes of the parents.  Each index of the offspring's binary string is the same as the same index of one of their parents, chosen at random.  The offspring then undergo random mutations, where some genes are randomly altered in order to reintroduce new features and diversity into the population.  The offspring become the new population and the cycle continues. 

This algorithm was tested on two very simple functions.  Each function had two sections of \enquote{bad} data, one where the outputs were too large, and one where the outputs were too small, but in one function, all the \enquote{bad} data was simply shifted by a certain constant, and in the other, the \enquote{bad} data was noisy.  This genetic algorithm was much more successful at excluding the noisy data -- which makes sense because getting rid of one noisy bad data point would always reduce the lengthscale of the model, but getting rid of one non-noisy bad datapoint would not, since the model still had to fit to all the other bad datapoints.  This led this algorithm to find a local optimum of excluding all points near the bad section instead.  

\begin{figure}[h!]
\centering
\begin{subfigure}{.5\textwidth}
  \centering
  \includegraphics[width=.7\linewidth]{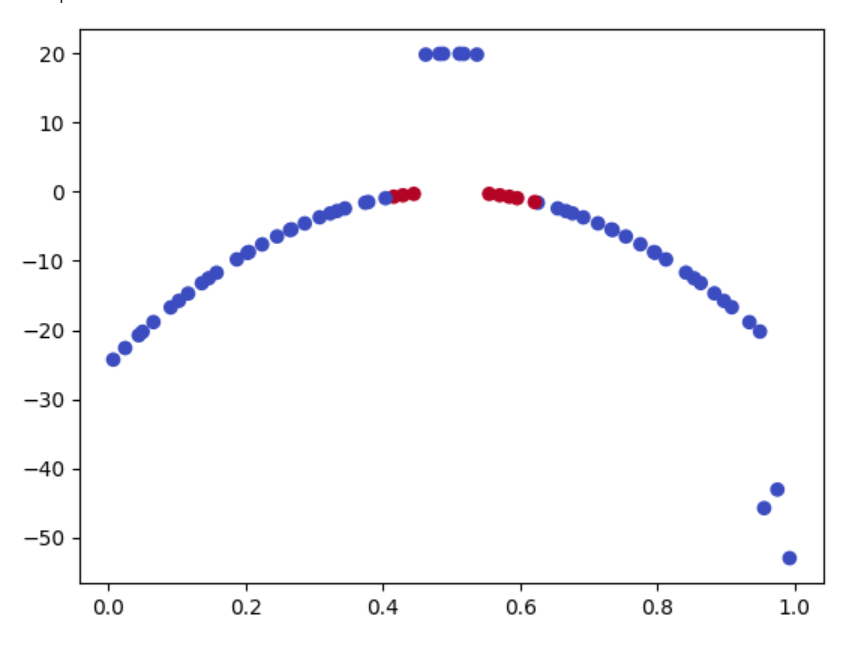}
  \caption{Test function with no noise on bad data}
  \label{fig:sub1}
\end{subfigure}%
\begin{subfigure}{.5\textwidth}
  \centering
  \includegraphics[width=.7\linewidth]{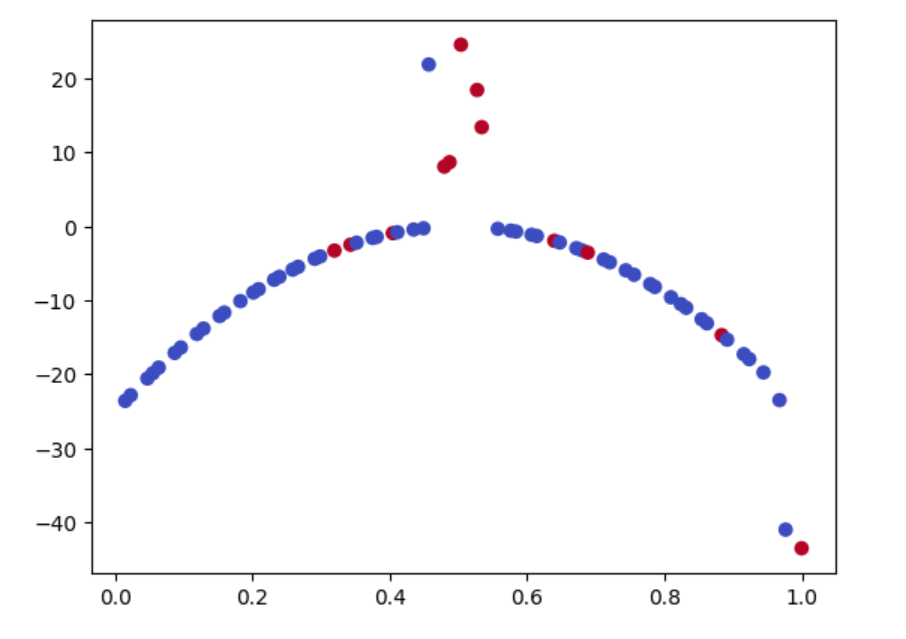}
  \caption{Same test function, with noise on bad data}
  \label{fig:sub2}
\end{subfigure}
\caption{Lengthscale-based genetic algorithm classification of which data points to exclude, shown in red}
\label{fig:test}
\end{figure}

These results prove promising for exclusion of faulty data points due to noise, which are common in the beam image data set.  When tested on the simulated beam function alongside the other pruning method, the genetic algorithm yielded similar results, though its runtime was significantly longer.  Further explorations of this algorithm for data pruning may include a restriction on model noise which may yield artificially large lengthscales, and investigations of intelligent ways to generate an initial population.  This would function as a prior to our algorithm, leading to even faster runtimes and more accurate models, potentially enabling us to dynamically alter the trust domains of the agent.  

\section{Conclusions}

\noindent This study has been a successful investigation into methods of data exclusion for optimal model training.  The dynamic pruning method, utilized with models trained on multiple beam features, proved effective in distinguishing between high and low-quality data. By dynamically excluding points that deviated significantly from model predictions, we achieved faster convergence to optimal beam configurations. This method was particularly adept at handling the complexities of noisy beam image data, ensuring the optimization process remained focused on high-fidelity readings.  As a result, a pruning method which applies this algorithm has been developed and integrated into the beamline alignment code at NSLS-II.  

These research approaches co-align with the overarching goal of Brookhaven National Laboratory to advance scientific research by providing consistently high-quality beams. The integration of these techniques into the beamline alignment code at NSLS-II marks a significant step towards achieving autonomous, reliable, and efficient beamline alignment.  Though these successes are tangible, this study is far from complete.  Future work will focus on refining these techniques, exploring their applicability across different beamlines, and investigating intelligent ways to further reduce runtime. These advancements will not only improve the efficiency and reliability of beamline alignment but also enhance the reproducibility of experiments and the overall scientific output at NSLS-II and other synchrotron facilities.

\section{Acknowledgement}

\noindent This project was supported in part by the U.S. Department of Energy, Office of Science, Office of Workforce Development for Teachers and Scientists (WDTS) under the Science Undergraduate Laboratory Internships Program (SULI).  

\section{References}

$^{1}$ Morris, T. W., M. Rakitin, A. Islegen-Wojdyla, Y. Du, M. Fedurin, A. C. Giles, D. Leshchev et al. "A General Bayesian Algorithm for the Autonomous Alignment of Beamlines." \textit{https://arxiv.org/abs/2402.16716v1} (2024).

\end{document}